\documentclass[preprint,authoryear,12pt]{elsarticle}

\usepackage{epsfig}

\usepackage{amssymb}

\usepackage{hyperref}
\hypersetup{
    colorlinks=true,
    linkcolor=blue,
    filecolor=blue,      
    urlcolor=blue,
    citecolor=blue,
    filecolor=blue,
}

\journal{Advances in Space Research}

\begin{document}

%%%%%%%%%%%%%%%%%%%%%%%%%%%%%%%%%%%%%%%%%%%%%%%%%%%%%%%%%%%%%%%%%%%%%%%%%%%%%
%% Frontmatter
\begin{frontmatter}


% Use the tnoteref command within \title and fnref within \author or \address for footnotes;
% use the corref command within \author for corresponding author footnotes;
% \author{Name\corref{cor1}\fnref{label2}}
% \address{Address\fnref{label3}}

\title{A study of the composite supernova remnant MSH 15$-$5{\it 6} with \textit{Suzaku}}

\author[1]{Nergis Cesur\corref{cor}}
\address[1]{Department of Astrophysics/IMAPP, Radboud University, 6525 AJ Nijmegen, The Netherlands}
\cortext[cor]{Corresponding author}
\ead{N.Cesur@astro.ru.nl}
\author[2]{Aytap Sezer}
\address[2]{Department of Electrical-Electronics Engineering, Avrasya University, 61250, Trabzon, Turkey}
\ead{aytap.sezer@avrasya.edu.tr}
\author[3]{Jelle de Plaa}
\address[3]{SRON Netherlands Institute for Space Research, Sorbonnelaan 2, 3584 CA Utrecht, The Netherlands}
\ead{j.de.plaa@sron.nl}
\author[3,4,5]{Jacco Vink}
\address[4]{Anton Pannekoek Institute for Astronomy, University of Amsterdam, Science Park 904, 1098 XH Amsterdam, The Netherlands}
\address[5]{GRAPPA, University of Amsterdam, Science Park 904, 1098 XH Amsterdam, The Netherlands}
\ead{j.vink@uva.nl}

\begin{abstract}

The Galactic supernova remnant (SNR) MSH 15$-$5{\it 6} is a member of the class of composite SNRs that consists of the remnant shell and a displaced pulsar wind nebula (PWN). The earlier X-ray observations reported the comet-like morphology of the PWN and the ejecta distribution of the SNR. In this work, we present a study of MSH 15$-$5{\it 6} using archival \textit{Suzaku} data. We investigate the nature of the emission and spectral parameters of the remnant. The X-ray spectra are well fitted with a combination of a thermal and non-thermal model with temperature $\sim$0.6 keV and photon index $\sim$2.0. The slightly enhanced abundances of Ne, Mg, S and enhanced abundance of Si confirm the presence of ejected material.

\end{abstract}


\end{frontmatter}

\parindent=0.5 cm

%%%%%%%%%%%%%%%%%%%%%%%%%%%%%%%%%%%%%%%%%%%%%%%%%%%%%%%%%%%%%%%%%%%%%%%%%%%%%
%% Main text
\section{Introduction}
\label{intro}
Composite supernova remnants (SNRs) are characterized by a system containing an expanding shell into a surrounding medium and a pulsar wind nebula (PWN). The X-ray studies of composite SNRs provide detailed information about of these systems (e.g., MSH 11$-$62: \citealt{slane12}; Kes 75: \citealt{reynolds18}).

MSH 15$-$5{\it 6} (G326.3$-$1.8, Kes 25) is a member of the class of composite SNRs and has a PWN in a cometary morphology. The bow shock of the PWN runs supersonically through the interstellar medium (ISM). The SNR shell and the displaced plerionic component were found using high-resolution radio observations \citep{green97, dickel00}. The compact hard X-ray nebula at the Southwest (SW) rim of the SNR was discovered by \citet{plucinsky98}. The SNR has a circular shell structure in the radio band with a radius of $\sim$19$^\prime$ \citep{green14}. Its distance was assumed to be $\sim$4.1 kpc by a study based on H$\alpha$ radial velocity measurements \citep{rosado96}. In the radio band, it has been revealed that there is a plerion-like feature closer to the center of the SNR \citep{clark75, milne79}. In their \textit{XMM-Newton} and \textit{Chandra} study, \citet{yatsu13} detected a point source emitting a power-law spectrum with a photon index of $\sim$1.3, and they found spectral steepening between the spectral photon indices of $\sim$1.7 and $\sim$2.5 along the flow line from the apex toward the tail, and also an expanding shell at $\sim$860 km s$^{-1}$ in the Sedov phase. Beside of the steepening photon index with distance from the putative pulsar, \citet{temim13} also detected a thermal component with a temperature of $\sim$0.5 keV. Using the radio emission, \citet{swaluw01} estimated the ratio between the PWN radius and the SNR radius of MSH 15$-$5{\it 6} ($R_{\rm pwn}$/$R_{\rm snr}$=0.28) and derived the initial spin period ($P_{0}$) of the pulsar residing in the SNR to be 37 ms. Using two different epochs of the archival \textit{Chandra} data, \citet{temim17} found a pulsar velocity of $\sim$720 km s{$^{-1}$}, and its direction of motion is 14$^{\circ}$ $\pm$ 22$^{\circ}$ SW, and the ambient density increases from East to West.

In this work, we present the first results of the high spectral resolution {\it Suzaku} observation of the SW rim of MSH 15$-$5{\it 6}. Although the X-ray properties of the thermal and non-thermal emission were studied in detail by \citet{yatsu13} and \citet{temim13}, we study this SNR using more recent atomic database (AtomDB)\footnote{http://www.atomdb.org} version 3.0.9 and give a comparison of metal abundances with {\sc xspec}\footnote{https://heasarc.gsfc.nasa.gov/xanadu/xspec/} and {\sc spex}\footnote{https://www.sron.nl/astrophysics-spex} software packages, which provide almost consistent results with each other except a few parameters. We also perform an {\it Athena} simulation to investigate the ejecta feature within this SNR. The paper is organized as follows. In Section \ref{obs and red}, we describe \textit{Suzaku} observation and data reduction. In Section \ref{analysis and result}, we present spectral analysis and results. In Section \ref{disc}, we discuss the nature of the X-ray emission and the spectral properties of the SNR. Finally, we present our conclusions in Section \ref{conc}.

%% Observation and Data Reduction

\section{Observation and data reduction}\label{obs and red}

We use 76.3 ks archival \textit{Suzaku} data of the SW region of the SNR MSH 15$-$5{\it 6} (Obs.ID: 507039010, PI: Y. Yatsu) that was observed on 2013 February 3. The observation was performed by the X-ray Imaging Spectrometer \citep[XIS;][]{koyama07}. The XIS instrument consists of four X-ray CCD cameras on the focal planes of the X-Ray Telescope \citep[XRT;][]{serlemitsos07}. The \textit{Suzaku} CCDs have a high spectral resolution and low background. The XIS0, 2 and 3 cameras have front-illuminated (FI) CCDs, where the XIS1 is back-illuminated (BI). XIS0, XIS1 and XIS3 are available in this observation. In order to show the location of the regions on the structure of the SNR, we also use a {\it ROSAT} PSPC image that was obtained with an exposure time of 3975 s on 1991 March 14. 

For data reduction and analysis, we used HEADAS software version 6.20 and {\sc xspec} version 12.9.1 \citep{arnaud96} with AtomDB 3.0.9 \citep{foster12}. Furthermore, spectral analysis was done using the {\sc spex} package \citep{kaastra96, kaastra18} version 3.04.00 with {\sc spexact} version 3.04.00. We used the {\texttt xisrmfgen} and {\texttt xissimarfgen} tools to create the redistribution matrix file and the ancillary response file, respectively \citep{ishisaki07}. All spectra were grouped with a minimum of 30 counts bin$^{-1}$.

%%%  ANALYSIS and Result

\section{Analysis and result}
\label{analysis and result}

\begin{figure}
\centering \vspace*{1pt}
\includegraphics[width=0.68\textwidth]{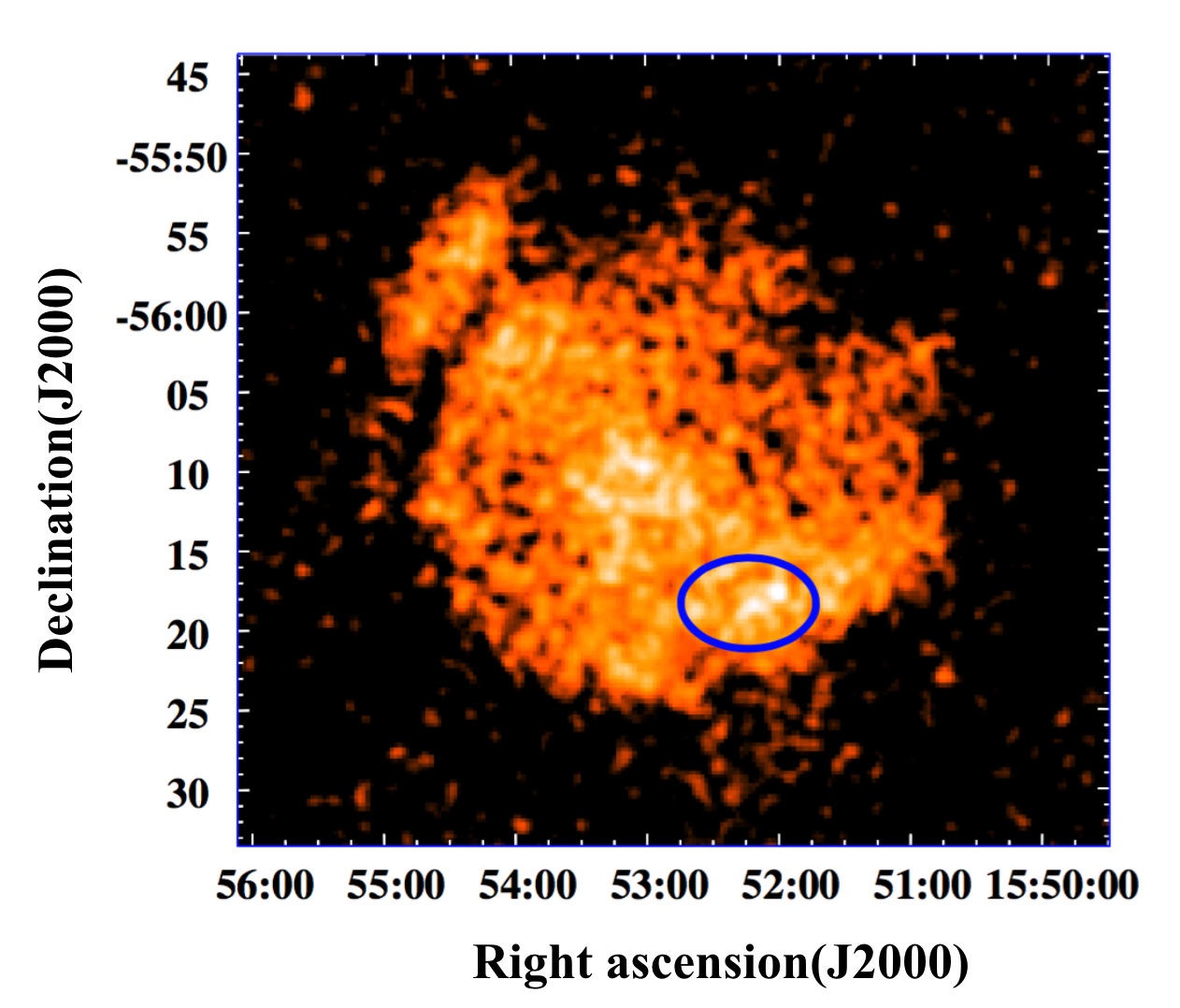}
\includegraphics[width=0.69\textwidth]{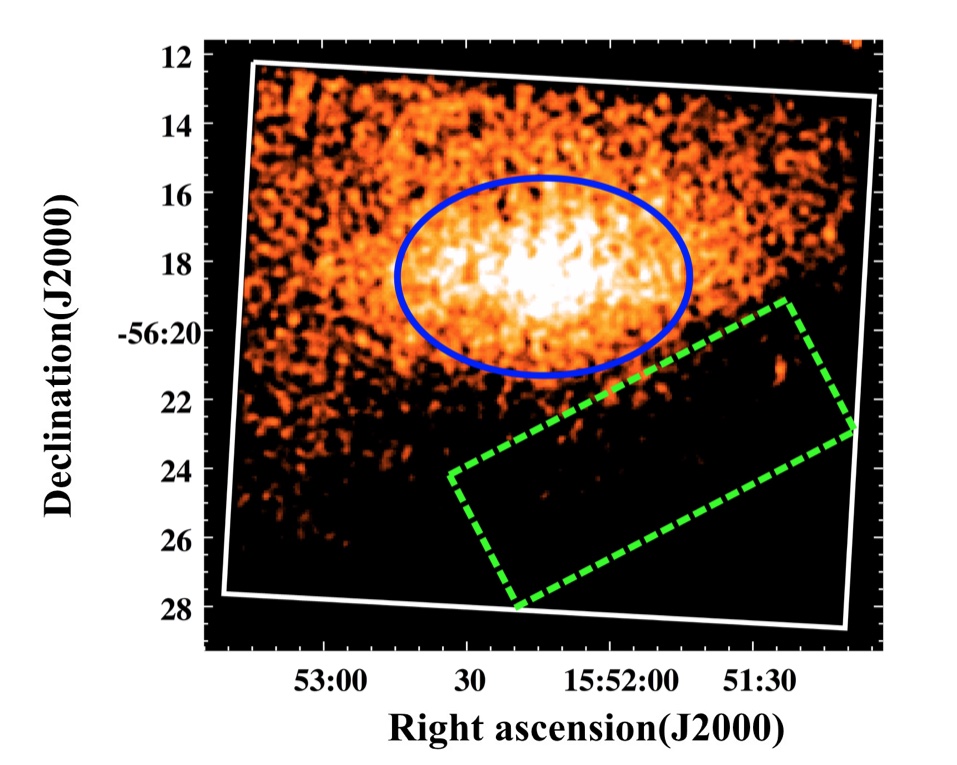}
\caption{Top: {\it ROSAT} PSPC image of MSH 15$-$5{\it 6} in the 0.1$-$2.4 keV energy band. The source region that contains SW region is shown by the blue elliptical area and utilized for the {\it Suzaku} spectral analysis. Bottom: {\it Suzaku} XIS1 image of the SW rim of MSH 15$-$5{\it 6} in the 0.3$-$10 keV energy band. The SW region is shown by the blue elliptical area. The background region is indicated by the dashed box, and the white box represents the FoV of XIS1. In both images, North is up and East is to left.}
\label{figure1}
\end{figure}

% X-ray Spectral Analysis
\subsection{X-ray image} \label{Spectral-analysis}

Figure \ref{figure1} (top) shows the {\it ROSAT} image of the SNR MSH 15$-$5{\it 6} in the 0.1$-$2.4 keV energy band. In this figure, the blue elliptical area represents the SW region that was selected for the {\it Suzaku} spectral analysis. Subsequently, Figure \ref{figure1} (bottom) shows the XIS1 image of this region in the 0.3$-$10 keV energy band. In order to characterize the emission, we extract the spectrum from an elliptical area with 4.2 arcmin $\times$ 2.9 arcmin centered at R.A.(J2000) = $15^{\rm h}52^{\rm m}13^{\rm s}.89$, Dec.(J2000)=$-56^{\circ}$18$'$28$''$.23, which is shown by blue ellipse on the figure. The background region is shown by the dashed box.

\subsection{Background estimation} 
The background spectrum was extracted from the region near the edge of the rim for the purposes of the spectral modeling (see Figure~\ref{figure1}, bottom). The average surface brightness of the background region is up to 23 per cent of those of the source region in the 0.6$-$7 keV energy band. The Non-X-ray Background (NXB) spectrum for the source and background was estimated from the night-earth data using {\texttt xisnxbgen} \citep{tawa08}. From the source and background spectrum, we subtracted the NXB spectrum using the {\sc mathpha} tool before the spectral fit.

% Spectral fitting
\subsection {Spectral fitting}

We carried out the spectral fitting in the source region both with {\sc xspec} and {\sc spex}. We first tried to fit the spectra with a single-component non-equilibrium ionization (NEI) model. The free parameters are the normalization, the absorption ($N_{\rm H}$), the electron temperature (\textit{kT$_{e}$}), the ionization time-scale ($\tau$ = $n_{\rm e}$t, where $n_{\rm e}$ and \textit{t} represent the mean electron density and the elapsed time after the shock heating of the plasma to a constant temperature, \textit{kT$_{e}$}), and the abundances of Ne, Mg, Si and S, respectively. Because of the statistically unacceptable fit ($\chi^{2}$/d.o.f $>$ 1.2), we added a {\texttt power-law} model to search for the non-thermal emission component and eventually obtained statistical improvement on the fits between the 0.6$-$7 keV energy range ($\chi^{2}$/d.o.f $\sim$ 1). 

Fitting the spectrum with this model, we were able to satisfactorily characterize the SW region with the following components: {\texttt power-law} plus {\sc vnei} model, both coupled with the same absorption model, which was set to vary. We used the absorption models of Tuebingen-Boulder ISM absorption model ({\sc tbabs}) \citep{wilms00} (in {\sc xspec}) and {\sc hot} (in {\sc spex}), (see {\sc spex} manual\footnote{https://www.sron.nl/astrophysics-spex/manual} and \citealt{kaastra09}) to account for the interstellar absorption from our Galaxy. To reckon with the neutral plasma gas limit, we fixed the temperature of the {\sc hot} model to 0.5 eV \citep{kaastra09}. According to the residuals and goodness-of-fit, we thawed the elemental abundances of \textsc{nei}, both in \textsc{xspec} (with {\sc nei} version 3.0) and {\sc spex}. Correlative to the best-fit parameters, the elemental abundances of Ne, Mg, Si and S are free parameters, while the other metal abundances were fixed to the solar values relative to the abundance tables of \citet{anders89} to compare between the \textsc{xspec} and \textsc{spex} fit directly, which is available in both spectral codes. 

\begin{figure}
\centering \vspace*{1pt}
\includegraphics[width=0.95\textwidth]{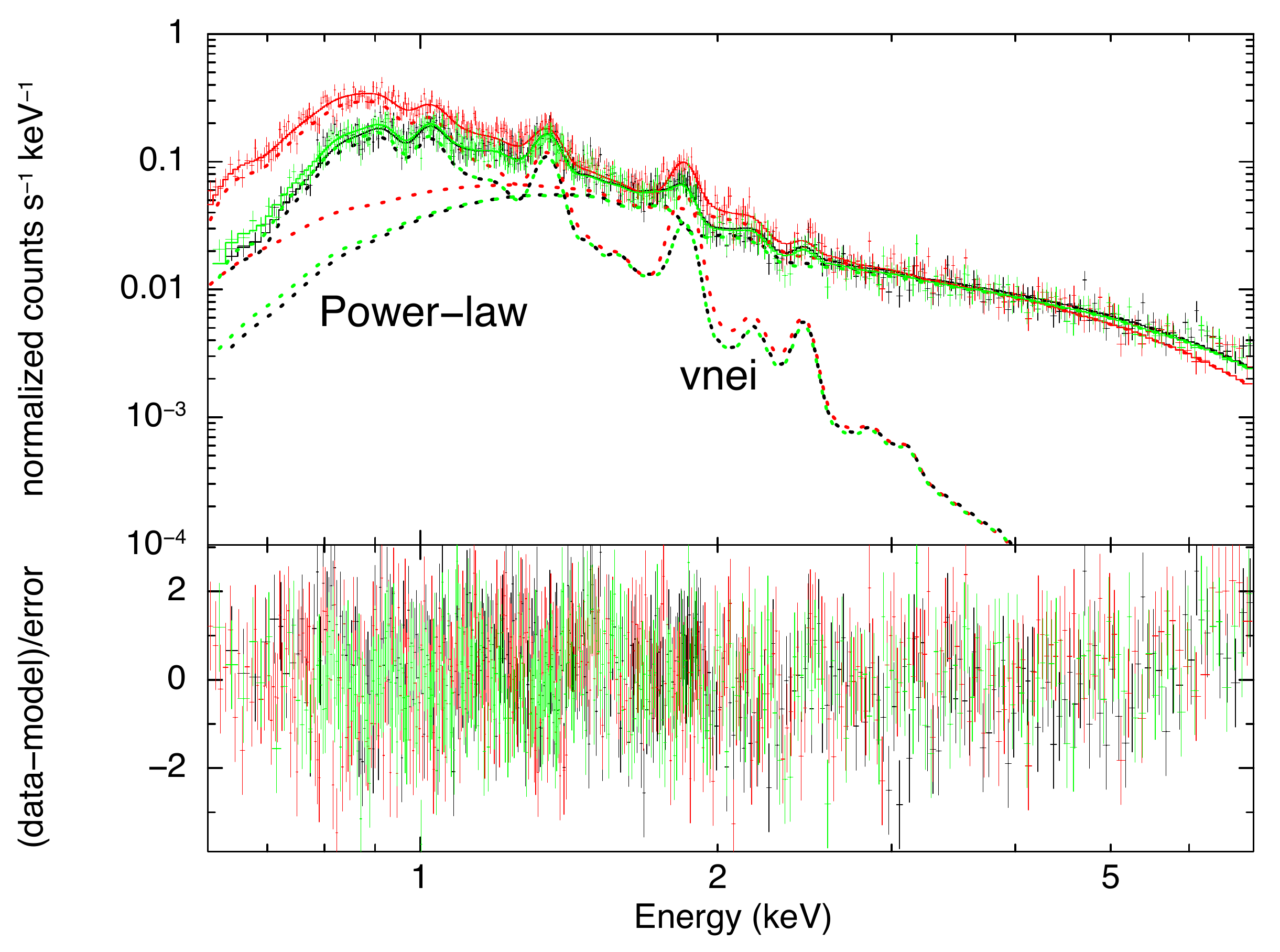}
\caption{The XIS spectra of the SW region of MSH 15$-$5{\it 6} in the 0.6$-$7 keV energy band fitted with the tbabs$\times$(vnei+power-law). The XIS0, 1 and 3 data are shown by crosses in black, red and green, respectively. The dotted lines represent vnei and power-law components. Residuals are plotted in the bottom panel.} 
\label{fig:xspec_spec}
\end{figure}

The best-fit spectral results are listed in Table~\ref{tab:table_xspec}, and the XIS spectra are displayed in Figure~\ref{fig:xspec_spec}. We calculated the uncertainties at the 90 per cent confidence level with the \texttt{error} command of \textsc{xspec} and \textsc{spex}.\\

\begin{table*}[t!]
\begin{center}
\caption{\label{tab:table_xspec}\textit{Suzaku} spectral fitting results with \textsc{xspec} and \textsc{spex}.}
\begin{tabular}{@{}p{2.2cm}p{3.3cm}p{2cm}p{2cm}@{}}
  \hline
  &&\multicolumn{2}{c}{Value}\\
    \cline{3-4}
Component           & Parameter                              & {\sc xspec}         &    {\sc spex}   \\
\hline \\[-2.0ex]
Absorption          & $N_{\rm H}$ ($10^{21}$ cm$^{-2})$      &  3.2$\pm{0.3}$      &  2.2$\pm{0.2}$ \\ [0.15 cm] 

NEI                 &  $kT_{\rm e}$ (keV)                    &  0.64$\pm{0.03}$    &  0.63 $\pm{0.01}$\\ [0.15 cm] 

                    &  Ne (solar)                            &  1.6$\pm{0.2}$      &  1.3$\pm{0.1}$ \\ [0.15 cm]
                    
                    &  Mg (solar)                            &   1.7$\pm{0.2}$     & 1.8$\pm{0.1}$\\ [0.15 cm]

                    &  Si (solar)                            &   2.2$\pm{0.4}$     & 2.3$\pm{0.2}$\\ [0.15 cm]
                   
                    &  S (solar)                             &   1.7$\pm{0.4}$     & 1.6$\pm{0.5}$\\ [0.15 cm]

                    &  $n_{\rm e}$t ($10^{11}$ cm$^{-3}$ s)  &  1.4$\pm{0.2}$      &  2.0$\pm{0.1}$\\ [0.15 cm]

                    & norm$^{\dagger}$ ($10^{-3}$)           &  1.0$\pm{0.2}$      &  ... \\ [0.15 cm]
                    & norm$^{\ddagger}$                      &          ...       &  1.1$\pm{0.1}$\\ [0.15 cm]

Power-law           &  $\Gamma$                              &   2.04$\pm{0.04}$   & 2.02$\pm{0.02}$\\ [0.15 cm]

                    &  norm$^{\mid}$  ($10^{-4}$)            &  7.8$\pm{0.4}$      &  ... \\ [0.15 cm]
                    &  norm$^{\parallel}$                    &          ...       &  9.9$\pm{0.3}$\\ [0.15 cm]

                    & $\chi^{2}$ (d.o.f.)            	     &   1.06 (1119)      &  1.03 (409)\\ [0.15 cm]

 \hline
\end{tabular}
\end{center}
\begin{flushleft}
\item {\footnotesize {{\bf Notes.} Abundances are given relative to the solar values in the abundance table of \citet{anders89}.\\
\item $^{\dagger}$ The normalization (norm) of the thermal model (\textsc{vnei} in \textsc{xspec}) is derived by $10^{-14}$ $\int$ $n_{\rm e}$ $n_{\rm H}$ dV$/(4\pi d^{2})$ in units of $10^{-3}$ cm$^{-5}$, where $n_{\rm e}$ and $n_{\rm H}$ are the electron and Hydrogen densities and V is the volume of the source. \\
\item $^{\ddagger}$ The norm of the thermal model (\textsc{nei} in \textsc{spex}) is the emission measure (Y $\equiv$ $n_{\rm e}$ $n_{\rm H}$ V) in units of 10$^{66}$ m$^{-3}$.\\
\item $^{\mid}$ The norm of the power-law (\textsc{xspec}) is in units of photons cm$^{-2}$ s$^{-1}$ keV$^{-1}$ at 1 keV. \\
\item $^{\parallel}$ The norm of the power-law (\textsc{spex}) is in units of $10^{45}$ ph s $^{-1}$ keV$^{-1}$.}} 
\end{flushleft}

\end{table*}

%%% DISCUSSION

\section{Discussion}\label{disc}

\subsection{Thermal emission}

We detected thermal and non-thermal emission from the SW rim with \textit{Suzaku} data, which indicates the interactive relation between the PWN and the thermal emission, pointing out the material in the remnant interior and an interaction of the SNR reverse shock with the PWN \citep{gaensler03,borkowski16}. It is therefore likely that the thermal component integrated with the PWN spectrum may be explained by the morphological relation between the PWN and the shell region. This component dominates nearly half of the X-ray spectrum, accounting for $\sim$54 per cent of the total unabsorbed flux [$F_{\rm (thermal)}$=(3.85$\pm{0.74}$) $\times$ 10$^{-12}$ ergs cm$^{-2}$ s$^{-1}$]. Besides, we found slightly enhanced abundances of Ne, Mg, S and enhanced abundance of Si, which presence of an ejecta feature and confirm the earlier studies. In addition, the abundance table of \citet{anders89} which we used during spectral fitting, is an earlier table, but the difference with the recent solar abundances \citep{lodders09} is at maximum about 10 per cent for our fitted abundances and is comparable or smaller to the size of the error bars. Thus, impact of using this table on our fitting is not that high.

The \textit{Suzaku} X-ray spectrum was well described by the electron temperature of $kT_{\rm e}$ $\sim$ 0.64 keV, and with an ionization time-scale of $\tau_{\rm }$ $\sim$ 1.4 $\times$ $10^{11}$ cm$^{-3}$ s, which indicates that the plasma has not yet reached ionization equilibrium. The previous study of \citet{yatsu13} has shown that the thermal component in the SW rim has an electron temperature of $kT_{\rm e}$ = 0.61 (0.58$-$0.64) keV, which is consistent with our result [$kT_{\rm e}$ = 0.64 (0.61$-$0.67) keV]. Furthermore, compared with the results those of \citet{yatsu13}, we found higher elemental abundances of Ne, Mg and Si, and the ionization parameter, $n_{e}$t, is smaller than their result. A note of caution is due here since they extracted spectra from two regions in the same SW rim area, we only selected a single region which is covering them all. Therefore, the difference between the results might be explained by this factor. \citet{temim13} also found thermal emission parameters in their spectral regions [$kT_{\rm e}$ = 0.62 (0.42$-$0.92) keV, for \textit{region b} (\textit{trail}), \textit{XMM-Newton}; $kT_{\rm e}$ = 0.56 (0.54$-$0.57) keV, for region \textit{region c}, \textit{Chandra}]. In our fittings, the abundance of Si and S are higher than the solar abundances and these results reflect those of \citet{temim13}. The comparison of our findings with those of other studies confirms the interrelation between PWN and ejecta material.

Based on the normalization of the {\sc vnei} model, we estimated the SNR parameters considering the distance of 4.1 kpc. We then calculated the volume of the SNR to be $V\sim 10.7\times10^{57}fd_{4.1}^{3}$ ${\rm cm^{3}}$, where $f$ is the volume filling factor ($0<f<1$) and $d_{4.1}$ is the distance in units of 4.1 kpc. Assuming $n_{\rm e}=1.2n_{\rm H}$, we found an electron density of $n_{\rm e}$ $\sim$ 0.15$f^{-1/2}d_{4.1}^{-1/2}$ ${\rm cm}^{-3}$ and age of $\sim30f^{1/2}d_{4.1}^{1/2}$ kyr, suggesting that it is a middle-aged SNR. Derived age and electron density results corroborate the finding of \citet{yatsu13} and \citet{temim13}. We then calculated the total mass of the emitting gas, $M_{\rm X}$ $\sim 2 f^{1/2}d_{4.1}^{5/2}~{M_\odot}$ using $M_{\rm X}$=1.4$m_{\rm H}n_{\rm e}V$. The small X-ray-emitting mass indicates that the X-ray emission in the SW region arises from the ejecta.

 \subsection{Non-thermal emission}
 
The non-thermal component of the \textsc{xspec} and \textsc{spex} models yielded a photon index ($\Gamma$) values of 2.04$\pm{0.04}$ and 2.02$\pm{0.02}$, respectively. These findings are consistent with that of \citet{temim13} [$\Gamma$ = 2.01 (1.90$-$2.13), for \textit{region b} (\textit{trail}), \textit{XMM-Newton};  $\Gamma$ = 2.05 (1.97$-$2.16), for \textit{region c}, \textit{Chandra}].

The photon index values we found are also slightly higher than those of \citet{yatsu13} [$\Gamma$ = 1.87 (1.75$-$1.99)]. On the other hand, the estimated non-thermal X-ray spectrum of a typical PWN has a spectral photon index value in the range of 1.3$-$2.3 \citep{gotthelf02}, and our result is consistent with this value. From the X-ray emission in Figure~\ref{fig:xspec_spec}, it is apparent that there is a hard X-ray continuum above 2 keV, and in the study of \citet{yatsu13}, it is attributed to the presence of the arc and PWN components.

\subsection{{\it Athena} X-IFU simulation}

The future X-ray mission {\it Athena} \citep{nandra2013} may be able to reveal the nature of the ejecta of MSH 15$-$5{\it 6}. With this aim, in this subsection we simulated the {\it Athena} X-ray Integral Field Unit (X-IFU) \citep{barret2016, pajot2018} 50-ks observation using {\sc xspec} and the latest X-IFU response files. Figure 3 shows {\it Athena} X-IFU spectrum fitted with an absorbed {\sc vnei}+{\texttt power-law} model. Our simulation demonstrates the potential of X-IFU for revealing the thermal emission from MSH 15$-$5{\it 6}. Spectroscopy with X-IFU ($\sim$2.5 eV energy resolution) would help us separate the thermal and non-thermal components, and constrain the properties of SNR and progenitor.
 \newpage
 
\begin{figure}
\centering \vspace*{1pt}
\includegraphics[width=0.7\textwidth]{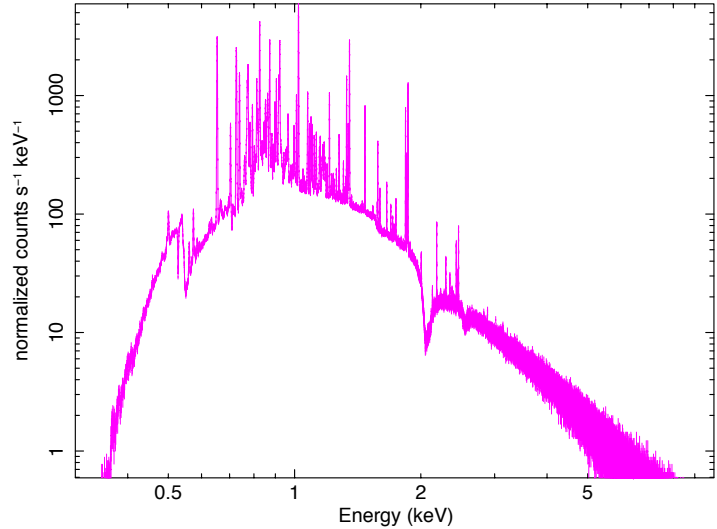}
\caption{50-ks {\it Athena} X-IFU spectrum (an absorbed {\sc vnei}+{\texttt power-law} model) of the SW region of MSH 15$-$5{\it 6} in the 0.3$-$10 keV energy band.}
\label{figure3}
\end{figure}

%%% CONCLUSION

\section{Conclusions}
\label{conc}

We presented a study of the SW region of the composite SNR MSH 15$-$5{\it 6}, utilizing the archival \textit{Suzaku} data. We described our main findings as follows.

\begin{enumerate}
\item The X-ray spectrum is well described by a two-component model with an absorption of $N_{\rm H}$  $\sim$ 3.2 $\times$ $10^{21}$ cm$^{-2}$: A non-thermal power-law model with a photon index of $\sim$2.04, and a thermal \textsc{nei} model with an electron temperature, $kT_{\rm e}$ $\sim$ 0.64 keV, an ionization time-scale, $\tau_{\rm }$ $\sim$ 1.4 $\times$ $10^{11}$ cm$^{-3}$ s.

\item The slightly enhanced abundances of Ne, Mg, S and enhanced abundance of Si support evidence of ejecta heated by the reverse shock. This result together with the small X-ray-emitting mass suggests that its emission arises from the shock heated ejecta.

\item Future X-ray observations with a high-resolution spectrometer such as {\it Athena} will allow us to examine the properties of thermal emission and the ejecta nature of MSH 15$-$5{\it 6}.

\end{enumerate}

\section{Acknowledgement}
The research leading to these results has received funding from the European Union's Horizon 2020 Program under the AHEAD project (grant agreement 654215).

\end{document}